\documentclass[runningheads]{svmult}

\usepackage{makeidx}   
\usepackage{graphicx}  
\usepackage{subeqnar}  
\usepackage{multicol}  
\usepackage{physprbb}  
\makeindex             

\newcommand{\MsunMarconi}{\ensuremath{~\mathrm{M}_\odot}}
\newcommand{\YRMarconi}{\ensuremath{\mathrm{\,yr}}}
\newcommand{\tenMarconi}[1]{\ensuremath{10^{#1}}}
\newcommand{\xtenMarconi}[1]{\ensuremath{\times 10^{#1}}}
\newcommand{\MBHMarconi}{\ensuremath{M_\mathrm{BH}}}
\newcommand{\rhoBH}{\ensuremath{\rho_\mathrm{BH}}}
\newcommand{\rhoBHunits}{\ensuremath{\,\times 10^5\MsunMarconi\mathrm{\,Mpc}^{-3}}}
\newcommand{\eff}{\ensuremath{\varepsilon}}
\newcommand{\MBHsigMarconi}{\MBHMarconi-\ensuremath{\sigma_\star}}
\newcommand{\MBHlumMarconi}{\MBHMarconi-\ensuremath{L_\mathrm{bul}}}

\begin{document}
\title*{Local Supermassive Black Holes,
Relics of Active Galactic Nuclei 
and the X-ray background}
\toctitle{Local Supermassive Black Holes, Relics of Active Galactic Nuclei 
and the X-ray background}
%
%
\titlerunning{Local Black Holes and AGN Relics}
%
\author{Alessandro Marconi\inst{1}
\and Guido Risaliti\inst{1,2}
\and Roberto Gilli\inst{1}
\and Leslie K. Hunt\inst{3}
\and Roberto Maiolino\inst{1}
\and Marco Salvati\inst{1}}
\authorrunning{Alessandro Marconi et al.}
%
%
\institute{INAF-Osservatorio Astrofisico di Arcetri,
Largo Fermi 5, I-50125 Firenze, Italy
\and Harvard-Smithsonian Center for Astrophysics,
60 Garden street, Cambridge, MA 02138, USA
\and INAF-Istituto di Radioastronomia-Sez.~Firenze,
	      Largo Fermi 5, I-50125 Firenze, Italy}

\maketitle              

\begin{abstract}
We summarize a study where we test the hypothesis that local
black holes (BH) are relics of AGN activity.  We compare the mass function of
BHs in the local universe with that expected from AGN relics, which are
BHs grown entirely with mass accretion during AGN phases. The local BH
mass function (BHMF) is estimated by applying the well-known correlations
between BH mass, bulge luminosity and stellar velocity dispersion to galaxy
luminosity and velocity functions. The density of BHs in the local universe is
$\rhoBH = 4.6_{-1.4}^{+1.9}\, h_{0.7}^2 \rhoBHunits$. The relic BHMF is derived
from the continuity equation with the only assumption that AGN activity is due
to accretion onto massive BHs and that merging is not important. We find that
the relic BHMF at $z=0$ is generated mainly at $z<3$. Moreover, the BH growth
is anti-hierarchical in the sense that smaller BHs ($\MBHMarconi<10^{7}M_\odot$)
grow at lower redshifts ($z<1$) with respect to more massive ones ($z\sim
1-3$).  Unlike previous work, we find that the BHMF of AGN relics is perfectly
consistent with the local BHMF indicating the local BHs were mainly grown
during AGN activity. This agreement is obtained while satisfying, at the same
time, the constraints imposed by the X-ray background. The comparison with the
local BHMF also suggests that the merging process is not important in shaping
the relic BHMF, at least at low redshifts ($z<3$). Our analysis thus suggests
the following scenario: local BHs grew during AGN phases in which
accreting matter was converted into radiation with efficiencies $\eff =
0.04-0.16$ and emitted at a fraction $\lambda = 0.1-1.7$ of the Eddington
luminosity. The average total lifetime of these active phases ranges from
$\simeq 4.5\times 10^{8}$ yr for $\MBHMarconi<10^{7}M_\odot$ to $\simeq 1.5\xtenMarconi{8}$
yr for $\MBHMarconi>10^{9}M_\odot$.
\end{abstract}

\section{Introduction}

The standard paradigm for Active Galactic Nuclei is that they are powered by
mass accretion onto a massive BH ($\MBHMarconi\sim \tenMarconi{6}-\tenMarconi{10}\MsunMarconi$). Combined
with the observed evolution of AGN, this implies that many (if not all) nearby
galaxies should host a BH in their nuclei as relic of past AGN activity.  BHs
are detected in $\sim 40$ galaxies and their mass correlates with host galaxy
structural parameters like bulge luminosity/mass \cite{kr95,MH03} and stellar
velocity dispersion \cite{fm00,gebhardt}.  An important open question is if
local BHs are relics of AGN activity (i.e.\ grown entirely with mass accretion
during AGN phases) or if other processes, such as merging, play an important role.
This can be answered by comparing the BHMF of local BHs with that expected from
AGN relics \cite{MS02,yu02}.  In recent work \cite{yu02,ferrarese} a
discrepancy in the BHMF at high masses ($\MBHMarconi>\tenMarconi{8}\MsunMarconi$) has been found:
more AGN relics are expected than
predicted by the local BHMF. This discrepancy can be reconciled
by assuming accretion efficiencies larger than the canonically adopted value of
$\eff=0.1$, i.e.\ $\eff>0.2$.  A refinement of the analysis by \cite{yu02} is
also presented in \cite{yu04} who find $\eff>0.1$.  High efficiencies are also
required from the comparison of $\rhoBH$ derived from the X-ray Background
(XRB) and from local BHs \cite{erz02}. Such high efficiencies, if confirmed,
would imply that most, if not all BHs, should be rapidly rotating.

In this paper we investigate the possibility that massive black holes in nearby
galaxies are relics of AGN activity by comparing the local BHMF with that of
AGN relics. The only assumption is that AGN activity is caused by mass
accretion onto the central BH.  The work is described in detail in
\cite{fullpaper} and here we focus on a few among the more critical and
important issues. Similar analysis, reaching conclusions analogous to ours are
presented in \cite{merloni,shankar} and by the same authors in these
proceedings. 

\section{The Mass Function of Local Black Holes}
\begin{figure}[t]
\centering
\resizebox{0.495\linewidth}{!}{\includegraphics{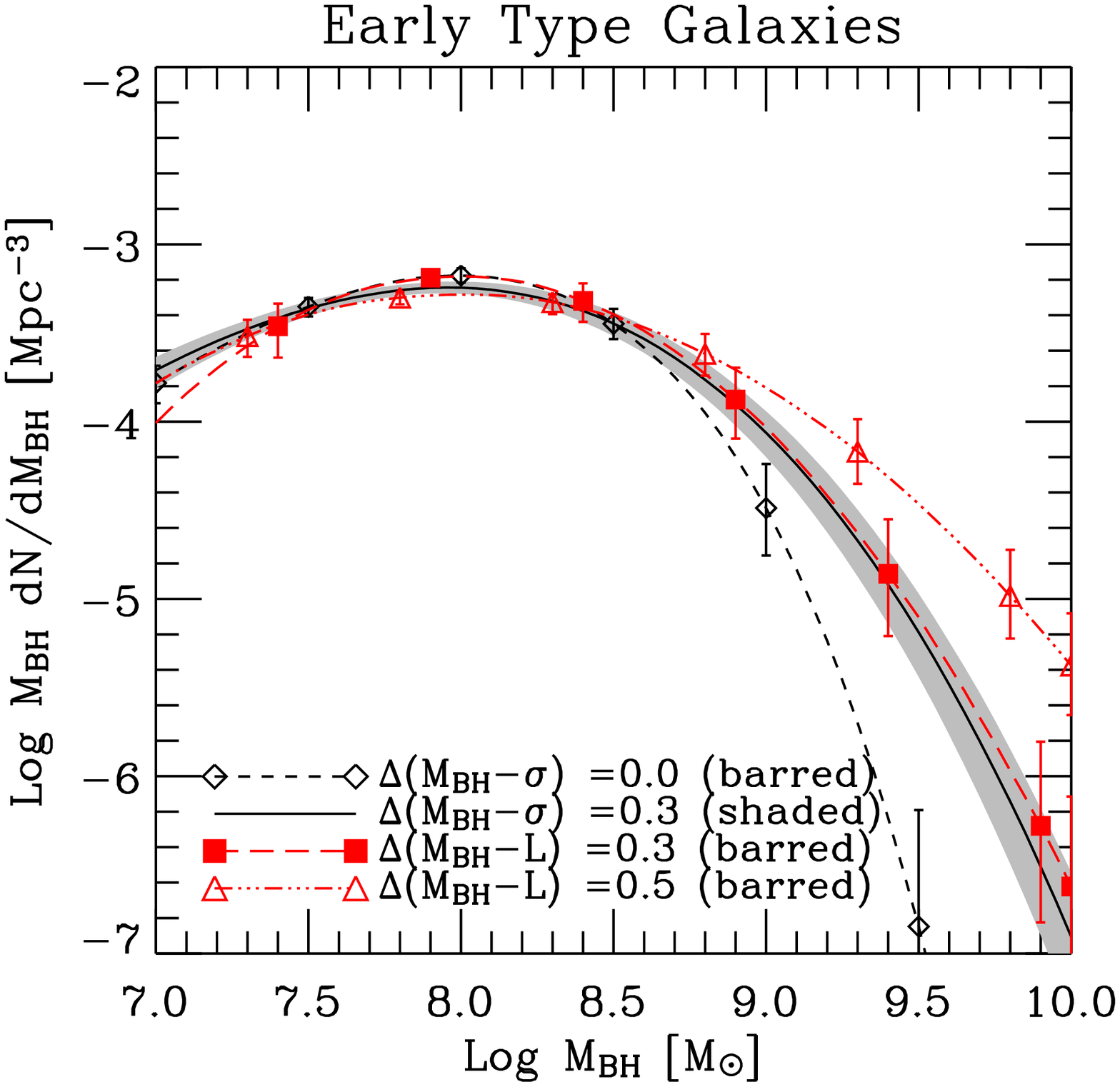} }
\resizebox{0.495\linewidth}{!}{\includegraphics{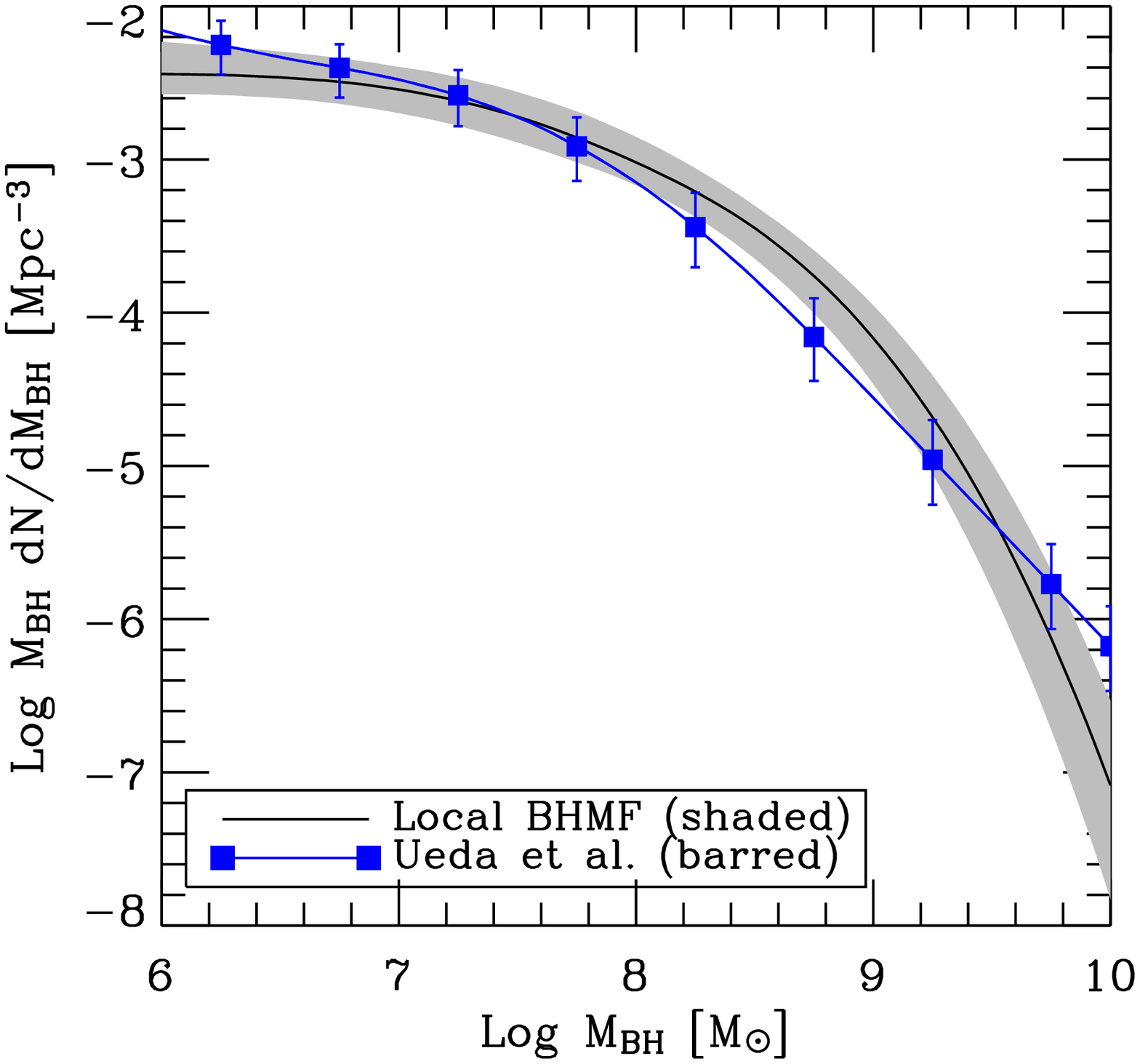} }
\caption[]{\label{fig:localBHMF}(a) Local BHMF for early type galaxies based on
the SDSS sample of \cite{bernardilum}. The shaded area and error bars
("barred") indicate 1$\sigma$ uncertainties. The $\Delta$ indicate the assumed
intrinsic dispersions of the \MBHsigMarconi\ or \MBHlumMarconi\ relations.  (b) Best estimate
of the local BHMF (shaded area) compared with the BHMF of AGN relics obtained
using the luminosity function by \cite{ueda}, corrected for the missing
Compton-thick AGNs.}
\end{figure}

The mass function of local BHs can be estimated by simply convolving the
existing galaxy luminosity [$\phi(L)$] or velocity functions [$\phi(\sigma)$]
with the \MBHlumMarconi\ and \MBHsigMarconi\ relations respectively. One should apply
corrections to convert from total to bulge luminosity in the first case and
should take into account the intrinsic dispersion (if any) of the \MBHMarconi-host
galaxy relations.  In Fig.\ \ref{fig:localBHMF}a we verify that the \MBHsigMarconi\
and  \MBHlumMarconi\ relations applied to the galaxy luminosity or velocity function
\cite{bernardilum,sheth} provide the same BHMF within the uncertainties
(estimated with many Montecarlo realizations of the BHMF). Clearly, the
necessary condition is that the two relations have the same intrinsic
dispersion since very different BHMFs are derived when \MBHlumMarconi\ has dispersion
0.5 in $\log\MBHMarconi$ at given $L_\mathrm{bul}$ and \MBHsigMarconi\ has 0 intrinsic dispersion.
This confirms the result by \cite{MH03} that all correlations \MBHMarconi--
host-galaxy-properties are equally good, i.e.\ they have similar intrinsic
dispersion. In Fig.\ \ref{fig:localBHMF}b  we plot the estimate of the local BH
mass function obtained considering galaxies from all morphological types. The
density in local BHs is $\rhoBH = 4.6 (-1.4;+1.9)(h/0.7)^2 \rhoBHunits$.  We
have used the galaxy luminosity functions by \cite{kochanek,nakamura,marzke}
and the galaxy velocity function by \cite{sheth}.

Our estimate of the local BH density is a factor $\sim 1.8$ larger than the
estimate by \cite{yu02}. \rhoBH\ is increased by $\sim 30\%$ when taking into
account an intrinsic dispersion for \MBHlumMarconi\ and \MBHsigMarconi\ (0.3 in $\log\MBHMarconi$ at
constant $\sigma_\star$ or $L_\mathrm{bul}$).  \rhoBH\ is also increased by $\sim 50\%$ because
we have used the zero points of the \MBHlumMarconi\ and \MBHsigMarconi\ correlations
determined by \cite{MH03}. These are a factor 1.5 larger than those used by
\cite{yu02} (see \cite{tremaine}) because they were determined by considering
only secure BH detections, where the BH sphere of influence is resolved by the
observations.  Fig.~8 of \cite{gebhardt03} show that measurements obtained with
data either resolving or not resolving the BH sphere of influence provide
similar \MBHMarconi\ values, albeit with much larger errorbars in the latter case.
However, from the same figure it can be evinced that on average BH measurements
where the sphere of influence is not resolved are underestimated by a factor
$\sim 2$, and this fully accounts for the larger zero points found by
\cite{MH03}. Apart from the larger zero points, by excluding the non-secure BH
detections \cite{MH03} find that \MBHsigMarconi\ and \MBHlumMarconi\ have the same dispersion
which, as we have just shown, is independently confirmed by the requirement of
obtaining the same BHMF from \MBHsigMarconi\ and \MBHlumMarconi\ applied to [$\phi(\sigma)$]
and [$\phi(L)$], respectively.

\section{The Mass Function of AGN Relics}

The Mass Function of AGN relics is estimated with the continuity equation which
relates the relic BHMF, $N(M,t)$, to the AGN luminosity function (LF),
$\phi(L,t)$ \cite{small}.  The only assumption is that AGNs are powered by mass
accretion onto a massive BH and that we can neglect merging of BHs. The
efficiency of mass-to-energy conversion is \eff\ and the BH is emitting at the
fraction $\lambda$ of the Eddington luminosity.  $L$ in the AGN LF must be the
'bolometric' luminosity. To obtain $L$, we derive and use bolometric
corrections which do not take into account the IR radiation (reprocessed UV
radiation). Thus, they are a factor $\sim 30\%$ lower than the values used by
previous authors.  We consider the Hard X-ray luminosity function by
\cite{ueda}, corrected for the missing Compton-thick AGNs (factor $\sim 1.6$)
and we apply a bolometric correction to obtain $\phi(L,t)$. Assuming that at
$z=3$ all BHs are active (this initial condition does not affect the final
results),  we can estimate the relic BHMF (with $\eff=0.1$ and $\lambda=1$) and
compare it with the local BHMF (Fig.\ \ref{fig:localBHMF}b).  The local BHMF
and the relic BHMF are in good agreement within the uncertainties. Thus, it is
unlikely that merging can play a major role in shaping the BHMF for $z<3$.  The
\cite{ueda} LF, corrected for the missing Compton-thick AGNs can also reproduce
the XRB spectrum and source counts, thus satisfying the constraints imposed by
the XRB.  In particular, the disagreement found by \cite{erz02} between the
density of local massive BHs and that inferred from the X-ray background light
can be reconciled noting that the average redshift of the sources making the
XRB is not $\langle z\rangle\simeq 2$ but $\langle z\rangle\simeq 1$, as shown
by the redshift evolution of the \cite{ueda} LF.
It is worth stressing the importance of the XRB constraint which effectively
removes one free parameter in this analysis, i.e.\ the fraction of obscured
AGNs. Clearly, a more refined fit of the XRB spectrum is
required and is the subject of future work.

\begin{figure}[t]
\centering
\parbox{0.495\linewidth}{\resizebox{\linewidth}{!}{\includegraphics{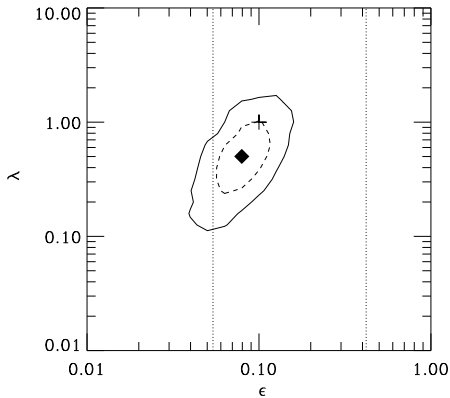}}}
\parbox{0.495\linewidth}{
\caption[]{\label{fig:efflambda} Locus where \eff\ and $\lambda$ provide the best match between local and relic BHMFs. The solid and dashed lines indicate an average deviation of 1 and 0.7$\sigma$ between the BHMFs. The diamond marks the \eff,$\lambda$ values providing the best agreement.}}
\end{figure}

\section{Accretion efficiency and $L/L_\mathrm{Edd}$}

Accretion efficiency \eff\ and Eddington ratio $\lambda$ are the only free
parameters for the relic BHMF and Fig.\ \ref{fig:efflambda} shows the locus
where they provide the best match between the relic and local BHMFs. The solid
and dashed lines shows the loci where the average deviation between the BHMFs
is less than 1 and 0.7$\sigma$, respectively. Outside of the solid contour, the
agreement between the BHMFs is poor. The dotted lines marks the \eff\ values
for a non-rotating Schwarzschild BH and a maximally rotating Kerr BH.
Acceptable values are in the range $\eff=0.04-0.16$ and $\lambda=0.1-1.7$.

The discrepancy found by \cite{yu02,ferrarese} is thus removed without
requiring large accretion efficiencies because of (i) the use of the zero
points of \cite{MH03}, (ii) the non-zero intrinsic dispersion of the \MBHsigMarconi\
and \MBHlumMarconi\ relations, and (iii) the smaller bolometric corrections.  While
(ii) and (iii) are generally accepted, (i) is more controversial because the
definition of `secure' BH mass measurement varies from author to author.
However, even using the same zero points as \cite{yu02} to estimate the local
BHMF we obtain $\eff=0.05-0.2$ and $\lambda=0.15-2.5$, i.e.\ efficiency is
still $\eff<0.2$.
It is worth noticing that the average limits on $L/L_\mathrm{Edd}$ that we find
are perfectly in agreement with the average values estimated by \cite{mclure04} on a large sample of SDSS quasars (see their Fig.~2).

\begin{figure}[t]
\centering
\resizebox{0.495\linewidth}{!}{\includegraphics{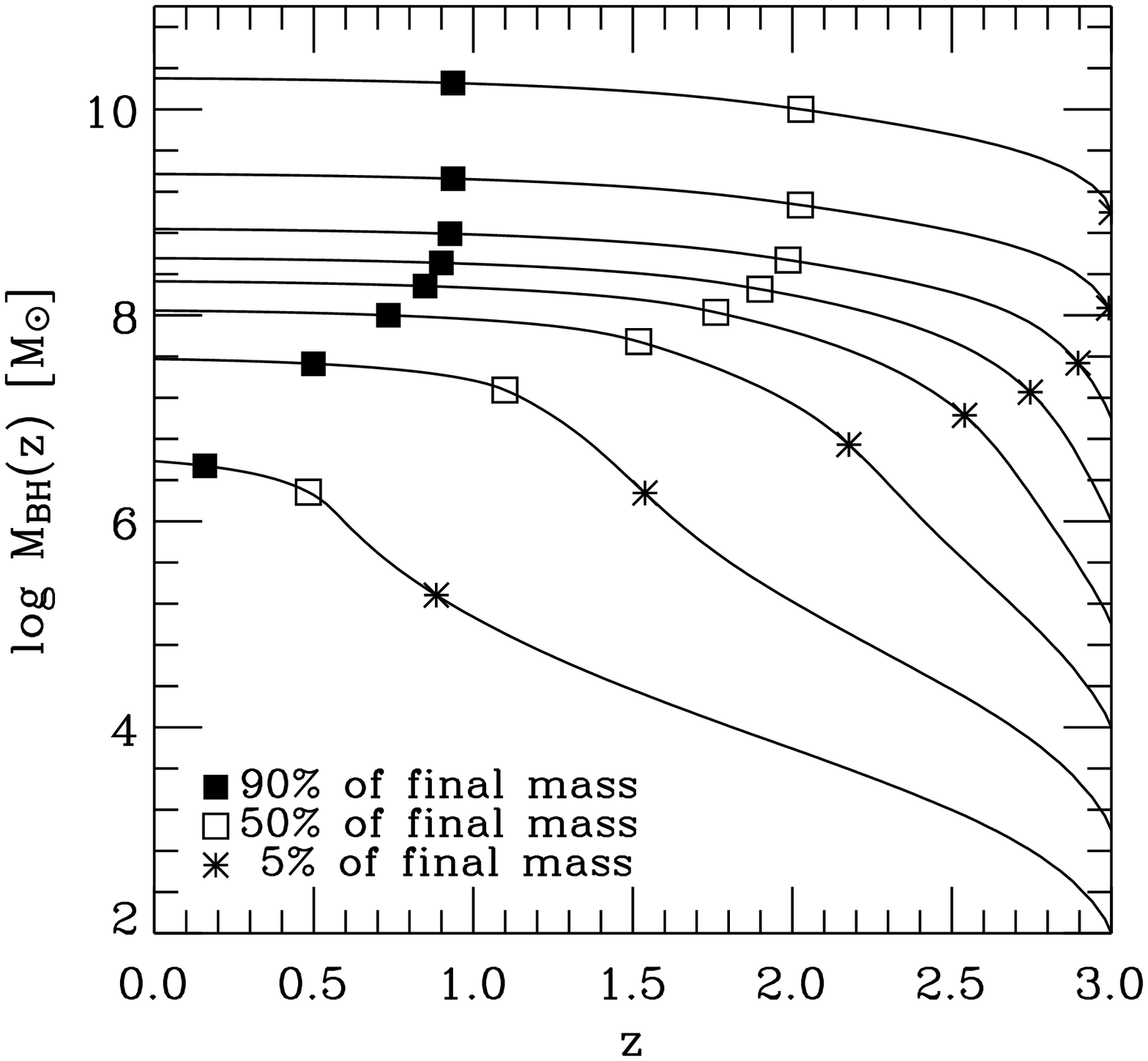} }
\resizebox{0.495\linewidth}{!}{\includegraphics{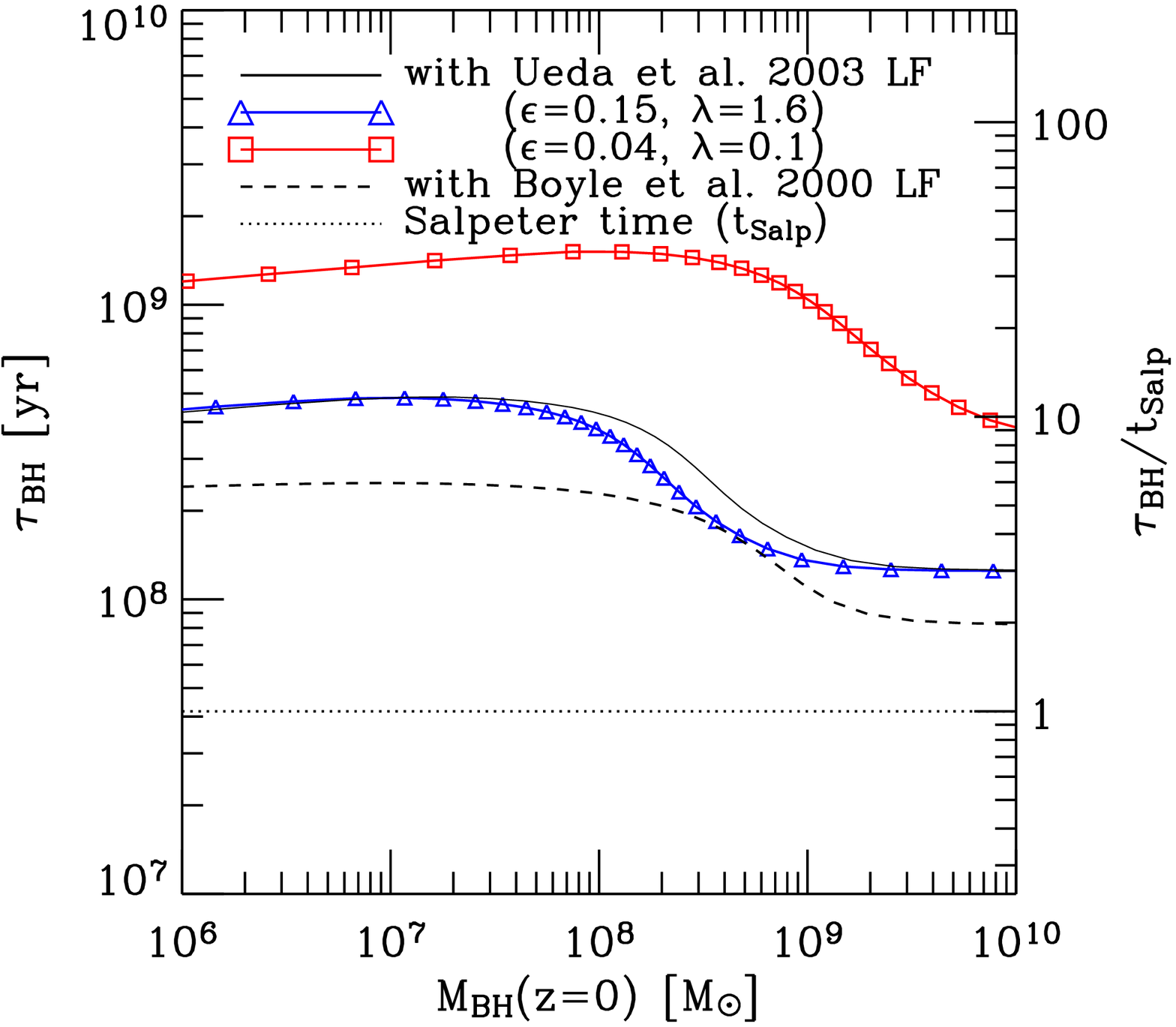} }
\caption[]{\label{fig:BHgrowth}(a) Average growth history of BHs. The symbols indicate the points when a BH reaches a given fraction of its final mass. (b) Average mean lifetimes of active BHs as a function of their mass at $z=0$. The solid line corresponds the standard case ($\eff=0.1$, $\lambda=1$). }
\end{figure}

\section{Anti-hierarchical BH growth and AGN lifetimes}

Fig.\ \ref{fig:BHgrowth}a shows the average growth history of BHs with
different starting masses at $z=3$. Symbols mark the point when a BH reaches a
given fraction of its final mass. At $z<3$, all BHs gain at least 95\% of their
final mass but BHs which are more massive than \tenMarconi{8}\MsunMarconi\ grow earlier and
gain 50\% of their final mass by $z\sim 2$. Smaller BHs grow at lower redshifts
($z<1$). This anti-hierarchical growth of BHs is a consequence of the redshift
evolution of the \cite{ueda} LF. 

Fig.\ \ref{fig:BHgrowth}b shows the average total lifetimes of active BHs,
i.e.\ the time required for the BH growth since $z=3$. The solid line shows the
"canonical" case with $\eff=0.1$, $\lambda=1$. Lines with symbols show limiting
cases from Fig.\ \ref{fig:efflambda}. Local high mass BHs ($\MBHMarconi<\tenMarconi{9}\MsunMarconi$)
have been active, on average, $\simeq 1.5\xtenMarconi{8}\YRMarconi$. On the contrary, the
assembly of lower mass BHs has required active phases lasting at least three
times that much ($\simeq 4.5\xtenMarconi{8}\YRMarconi$).  The average lifetimes can be as
large as \tenMarconi{9}\YRMarconi\ with the smaller \eff\ and $\lambda$ values compatible
with local BHs ($\eff=0.04$, $\lambda=0.1$ - see Fig.\ \ref{fig:efflambda}).

Overall, the plots in Fig.\ \ref{fig:BHgrowth} indicate that smaller BHs
($\MBHMarconi<\tenMarconi{8}\MsunMarconi$) find more difficulties in growing than larger ones.
Indeed, this is consistent with physical models for the coevolution of BHs and
galaxies. Smaller BHs form in shallower potential wells with respect to more
massive ones and are thus more subject to feedback from star formation (e.g.\
supernovae explosions) and from the AGN itself \cite{menci04}.

\section{Conclusions}

We have shown that the local BH mass function and that of AGN relics are in
good agreement with standard accretion efficiency and $L/L_\mathrm{Edd}$ ratio
($\eff\sim 0.1$, $\lambda\sim 1$).  In particular, the local BH Mass function
implies that the density in BHs is $\rhoBH = 4.6 (-1.4;+1.9)(h/0.7)^2
\rhoBHunits$, a factor 1.8 higher than estimated by \cite{yu02}.  Merging of BHs is
either not important or it does not significantly alter the relic BHMF, at
least at $z<3$.  The BH growth is anti-hierarchical, in the sense that smaller
BHs ($\MBHMarconi<\tenMarconi{7}\MsunMarconi$) grow at lower redshifts ($z<1$) with respect to more
massive ones ($z=1-3$).  The global picture which emerges is than that local
BHs grew during AGN phases in which accreting matter was converted into
radiation with $\eff=0.05-0.2$ and $L/L_\mathrm{bol}=0.15-2.5$. The average
total lifetime of these active phases ranges from $\simeq 4.5\xtenMarconi{8}\YRMarconi$
($\MBHMarconi<\tenMarconi{7}\MsunMarconi$) to $\simeq 1.5\xtenMarconi{8}\YRMarconi$ ($\MBHMarconi>\tenMarconi{9}\MsunMarconi$).

%

\end{document}